\documentclass[runningheads]{svmult}
\usepackage{makeidx}   
\usepackage{graphicx}  
\usepackage{subeqnar}  
\usepackage{multicol}  
\usepackage{cropmark} 
\usepackage{physprbb}  
\def\al{\alpha}
\def\be{\beta}
\def\ga{\gamma}
\def\de{\delta}

\def\ve{\varepsilon}

\def\et{\eta}
\def\th{\theta}

\def\la{\lambda}

\def\rh{\rho}

\def\si{\sigma}

\def\ta{\tau}

\def\De{\Delta}

\def\La{\Lambda}

\def\Om{\Omega}

\def\cl{{\cal L}}

\def\fr#1#2{{{#1} \over {#2}}}

\def\frac#1#2{{\textstyle{{#1}\over {#2}}}}

\def\lsim{\mathrel{\rlap{\lower4pt\hbox{\hskip1pt$\sim$}}
    \raise1pt\hbox{$<$}}}
\def\gsim{\mathrel{\rlap{\lower4pt\hbox{\hskip1pt$\sim$}}
    \raise1pt\hbox{$>$}}}
\def\sqr#1#2{{\vcenter{\vbox{\hrule height.#2pt
         \hbox{\vrule width.#2pt height#1pt \kern#1pt
         \vrule width.#2pt}
         \hrule height.#2pt}}}}

\def\prt{\partial}
\def\lrpartial{\raise 1pt\hbox{$\stackrel\leftrightarrow\partial$}}

\def\etal{{\it et al.}}

\newcommand{\beq}{\begin{equation}}
\newcommand{\eeq}{\end{equation}}
\newcommand{\bea}{\begin{eqnarray}}
\newcommand{\eea}{\end{eqnarray}}
\newcommand{\rf}[1]{(\ref{#1})}

\def\sech{\mathop{\rm sech}\nolimits}

\begin{document}
\title*{The Lorentz-violating extension\protect\newline
of the Standard Model}
\toctitle{The Lorentz-violating extension\protect\newline
of the Standard Model}
%
%
\titlerunning{The Lorentz-violating Standard-Model Extension}
%
\author{Ralf Lehnert}
\authorrunning{Ralf Lehnert}
%
%
\institute{CENTRA, Department of Physics, University of the Algarve, 
8000-117 Faro, Portugal}

\maketitle              

\begin{abstract}
Quantum-gravity effects are expected to be suppressed by the Planck mass. 
For experimental progress it is therefore important to identify potential 
signatures from Planck-scale physics that are amenable to ultrahigh-precision 
tests. It is argued that minuscule violations of Lorentz and CPT symmetry 
are candidate signals. In addition, theoretical and experimental aspects of 
the Standard-Model Extension, which describes the emergent low-energy 
effects, are discussed. 
\end{abstract}

\section{Motivation}

An important open question 
in our understanding of nature at its fundamental level 
concerns a unified quantum description 
of all fundamental interactions including gravity. 
Such a theory is likely to become dominant only 
as the Planck scale is approached, 
so that quantum-gravity effects 
are expected to be minuscule 
at presently attainable energies. 
Moreover, 
the absence of a fully realistic 
and viable candidate underlying theory 
provides a major obstacle 
for the identification of concrete quantum-gravity signatures 
that can be searched for 
in present-day or near-future experiments. 

A possible approach 
to overcome this phenomenological issue 
is to determine exact relations in the currently accepted laws of physics 
that may be violated in prospective fundamental theories 
and that can be tested with ultrahigh-precision. 
Symmetries typically satisfy these criteria. 
For example, 
Lorentz and CPT invariance 
are cornerstones of our present understanding of nature 
at the fundamental level, 
and a variety of Lorentz and CPT tests 
belong to the most sensitive null experiments available. 
Lorentz and CPT violation 
is also a key feature 
of certain approaches to underlying physics. 

For example, 
couplings varying on cosmological scales
are one possible source of Lorentz and CPT violation \cite{klp03}. 
This fact does not come as a surprise: 
parameters dependent on spacetime 
break translational invariance, 
and translations, rotations, and boosts 
are linked in the Poincar\'e group. 
Thus, 
violations of translation symmetry 
can also affect Lorentz invariance. 
This can be understood intuitively 
as follows. 
The equations of motion 
typically contain the gradient of the coupling, 
which selects a preferred direction in spacetime 
leading to apparent Lorentz violation. 

In this talk, 
we discuss some theoretical and experimental aspects 
of the Standard-Model Extension (SME) 
\cite{ck97,ck98,jk,kl01,klp02,grav}, 
which is the low-energy framework 
for Lorentz-breaking effects. 
The SME is a dynamical model  
constructed to contain all Lorentz- and CPT-violating 
lagrangian terms 
consistent with coordinate independence, 
which is a fundamental requirement 
to be discussed below. 
To date, 
numerous Lorentz- and CPT-violation tests 
involving hadrons 
\cite{ktev,opal,delphi,belle,focus,ck95a,ck95b,ck95c,kv96,baryo,k98,k99,k01,isg01}, 
protons and neutrons 
\cite{hu99,bear00,hum00,kl99a,kl99b,blu02,blu03,sudar02,phil01,pnthhh}, 
electrons 
\cite{pnthhh,deh99,mit99,gab99,bluhm97,bluhm98,bluhm00,heck02,iltan2,iltan3,rl04}, 
photons 
\cite{cfj,km,km02,brax,lipa,sh03}, 
muons \cite{muons1,muons2,iltan1}, 
and neutrinos \cite{ck97,cg99,neutrinos,neu1,neu2} 
have been analysed 
or identified 
within the framework of the SME. 

The outline of this talk is as follows. 
In Sec.\ \ref{ci}, 
we analyse the requirement of coordinate independence 
and its implementation. 
Section \ref{lvsme} contains a discussion 
of the SME. 
In particular, 
its construction is reviewed, 
its generality is addressed, 
and its advantages are summarized. 
In Sec.\ \ref{vc}, 
varying couplings are investigated 
from the perspective of providing a potential source 
of Lorentz and CPT violation. 
Section \ref{thres} 
comments on kinematical Lorentz tests 
with modified dispersion relations. 
The conclusions are contained in Sec.\ \ref{conc}. 

\section{Coordinate independence} 
\label{ci}

One of the most fundamental principles in physics 
is coordinate independence. 
The need for this principle 
in the presence of Lorentz breaking 
is well established \cite{ck97,ck98,rl03}, 
and it has served as the basis 
for the construction of the SME. 
However, 
in some investigations of Lorentz and CPT violation 
coordinate-{\it dependent} physics emerges, 
and occasionally Lorentz-symmetry breakdown is identified 
with the loss of coordinate independence. 
For these reasons, 
it is appropriate to review this fundamental principle
and its implementation. 
It then also becomes clear 
how coordinate independence 
provides a rough classification 
of different approaches to Lorentz and CPT breaking. 

A certain choice of labeling events in space and time 
is called a coordinate system. 
Such a labeling scheme 
is typically observer dependent 
and thus arbitrary to a large extent. 
Coordinate systems 
belong to the most common and important {\it tools} 
for the description of processes occurring in nature, 
but they fail to possess {\it physical reality}: 
the choice of coordinates 
must leave the physics unaffected. 
This principle of coordinate independence 
is fundamental in science. 
It assures 
that the physics remains independent of the observer, 
and it is therefore also called observer invariance. 
Coordinate independence is guaranteed 
when spacetime is given a manifold structure 
and physical quantities are represented 
by geometric objects, 
such as tensors or spinors. 

Coordinate-{\it dependent} physics 
does break Lorentz symmetry. 
However, 
the converse 
(i.e., Lorentz violation is associated 
with the loss of coordinate independence) 
is a common misconception. 
The principle of coordinate invariance is, 
in fact, 
independent from Lorentz symmetry. 
For instance, 
Newton's law of gravitation and nonrelativistic classical mechanics 
are non-invariant under Lorentz transformations 
but can be formulated in the coordinate-free language of 3-vectors. 
The Lorentz transformations acquire a significant role only 
on lorentzian spacetime manifolds 
where they implement changes between local inertial frames. 

Even on a lorentzian manifold, 
Lorentz symmetry may be broken. 
This fact can be illustrated 
in the conventional context 
of a classical point particle of mass $m$ and charge $q$ 
subjected to an external electromagnetic field $F^{\mu\nu}$. 
The equation of motion for such a particle reads
\beq 
m\:\fr{dv^{\mu}}{d\ta}=qF^{\mu\nu}v_{\nu}\; , 
\label{example} 
\eeq 
where $\ta$ is the particle's proper time 
and $v^{\mu}$ is its  4-velocity. 
Equation \rf{example} remains valid in all coordinate systems 
because it is a tensor equation. 
Thus, 
observer Lorentz symmetry is maintained. 
However, 
the external $F^{\mu\nu}$ background violates, 
for example, 
symmetry under arbitrary rotations of the charge's trajectory. 
Among the consequences of this noninvariance 
is the violation of angular-momentum conservation 
for the particle. 
Note the difference to coordinate changes, 
which leave unaffected the physics: 
here, 
only the trajectory is rotated, 
so that its orientation 
with respect to $F^{\mu\nu}$ 
can change. 
One then says that particle Lorentz symmetry is violated, 
despite the presence of observer invariance \cite{ck97,rl03}. 
It is important to point out 
that in the above example, 
the background $F^{\mu\nu}$ is a local electromagnetic field 
caused by other 4-currents 
that can in principle be controlled. 
Our external-field illustration 
therefore fails to contain Lorentz violation 
as a fundamental property of an effective vacuum. 

The above discussion suggests the possibility 
of classifying different approaches to Lorentz violation 
by their behavior under coordinate changes. 
In the remaining part of this section, 
we discuss such a classification. 

{\bf Models with coordinate-dependent physics.} 
Although it appears to be impossible 
to perform meaningful scientific investigations 
involving coordinate-dependent physics, 
such approaches to Lorentz breaking 
have been considered in the literature. 
More specifically, 
there have been two suggestions 
in the context of neutrino phenomenology: 
the first one forces the masses of particle and antiparticle 
to be different \cite{baren01}, 
while the second one attempts to build a model 
from positive-energy eigenspinors only \cite{baren02}. 
Both approaches are known 
to involve coordinate-dependent off-shell physics \cite{green02,green03}. 
In what follows, 
we do not consider these approaches further. 

{\bf Coordinate-covariant models involving non-lorentzian manifolds.} 
Another possibility to speculate 
how Lorentz invariance might be lost 
is the following. 
Local inertial frames have a structure 
different from the usual minkow{\-}skian one, 
so that Lorentz transformations 
no longer implement changes between inertial coordinates, 
i.e., 
observer Lorentz invariance is {\it replaced} 
by observer invariance under some other symmetry transformation. 
But nevertheless, coordinate independence is maintained. 
This point of view is taken 
in the so called ``doubly special relativities''
\cite{dsr1,dsr2}. 
We mention 
that both 
the viability and the physical interpretation of this approach 
appear to be controversial at the present time 
\cite{luk03,rem03,ahl02,gru03,schu03}. 
We leave 
such Lorentz-symmetry {\it deformations} 
unaddressed in the present work. 

{\bf Coordinate-independent models involving nontrivial vacua.} 
In this approach, 
a fully Lorentz-covariant underlying model 
generates a tensorial background 
resulting in apparent Lorentz violation. 
The basic idea 
parallels that of our above external-field example. 
However, 
the background is outside of experimental control 
and must be viewed as a property of the effective vacuum. 
Because of the lorentzian structure of the underlying manifold 
and the usual Lorentz-covariant dynamics at the fundamental level, 
this approach appears closest to established theories. 
The physical effects in such models 
are perhaps comparable to those inside certain crystals: 
the physics remains independent of the chosen coordinates, 
but particle propagation, 
for example, 
can be direction dependent. 
As an immediate consequence, 
one can locally still work 
with the metric $\et^{\mu\nu}={\rm diag}(1,-1,-1,-1)$, 
particle 4-momenta are still additive 
and still transform in the usual way under coordinate changes, 
and the conventional tensors and spinors 
still represent physical quantities. 

Such nontrivial effective vacua 
can arise in various theories beyond the Standard Model. 
For instance, 
the possibility 
of spontaneous Lorentz and CPT breaking 
in the framework of string field theory 
was discovered 
more than a decade ago 
\cite{ks89a,ks89b,ks89c,ks91,kp91,kp96,kpp00,kp01}. 
Subsequent studies 
have considered other mechanisms for Lorentz-violating vacua 
including spacetime foam \cite{ell98,suv}, 
nontrivial spacetime topology \cite{klink}, 
loop quantum gravity \cite{amu00,amu02}, 
realistic noncommutative field theories \cite{chklo,gur,ani,carl}, 
and spacetime-varying couplings \cite{klp03,blpr}. 

\section{The Standard-Model Extension}
\label{lvsme}

The next step after determining general low-energy features 
is the identification 
of specific experimental signatures for Lorentz breaking 
and the theoretical analysis of Lorentz-violation searches. 
This task is most conveniently performed 
within a suitable test model. 
Many Lorentz tests are motivated and analysed 
in purely kinematical frameworks 
allowing for small violations of Lorentz symmetry. 
Examples are  Robertson's framework, 
its Mansouri-Sexl extension, 
the $c^2$ model, 
and phenomenologically constructed modified dispersion relations. 
But is also clear 
that the implementation of general dynamical features 
significantly increases the scope 
of Lorentz tests. 
For this reason, 
the SME 
mentioned in the introduction 
has been developed. 
However, 
the use of dynamics in Lorentz-violation searches 
has recently been questioned 
on the grounds of framework dependence. 
We disagree with this claim  
and begin with a few arguments in favor of a dynamical test model. 

Such a model is constrained 
by the requirement 
that known physics must be recovered 
in certain limits, 
despite some freedom 
in introducing dynamical features 
compatible with a given set of kinematics rules. 
In addition, 
it seems difficult 
and may even be impossible 
to construct an effective theory containing the Standard Model 
with dynamics significantly different from that of the SME. 
We also mention 
that kinematical analyses 
are limited 
to only a subset of potential Lorentz-violating signatures 
from fundamental physics. 
From this viewpoint, 
it is desirable 
to explicitly implement dynamical features 
of sufficient generality 
into test models for Lorentz and CPT symmetry.  

{\bf The generality of the SME.}
In order to understand 
the generality of the SME, 
we review the main elements of its construction \cite{ck97,ck98}. 
Starting from the usual Standard-Model lagrangian ${\cal L}_{\rm SM}$ 
one adds Lorentz-violating modifications $\de {\cal L}$: 
\beq
{\cal L}_{\rm SME}={\cal L}_{\rm SM}+\de {\cal L}\; .
\label{sme}
\eeq
Here, the SME lagrangian is denoted by ${\cal L}_{\rm SME}$. 
The correction $\de {\cal L}$ 
is formed by contracting Standard-Model field operators 
of any dimensionality 
with Lorentz-breaking tensorial coefficients 
that describe the nontrivial vacuum 
discussed in the previous section. 
To ensure coordinate independence, 
this contraction must yield 
observer Lorentz scalars. 
It becomes thus apparent 
that all possible contributions to $\de {\cal L}$ 
determine the most general effective dynamical description 
of Lorentz violation 
at the level of observer Lorentz-invariant quantum field theory. 

Potential Planck-scale features, 
such as non-pointlike elementary particles 
or a discretized spacetime, 
are unlikely to invalidate 
the above effective-field-theory approach 
at presently attainable energies. 
On the contrary, 
the phenomenologically successful Standard Model 
is normally understood 
as an effective-field-theory approximation 
for more fundamental physics. 
If fundamental physics 
indeed exhibits minuscule Lorentz-breaking effects, 
it would seem contrived 
to consider low-energy effective models 
outside the framework of quantum field theory. 
We finally mention 
that the need for a low-energy description 
beyond effective field theory 
is also unlikely to arise 
in the context of candidate underlying models 
with novel Lorentz-invariant aspects, 
such as additional particles, 
new symmetries, 
or large extra dimensions. 
Lorentz-symmetric modifications 
can therefore be implemented into the SME, 
if necessary \cite{bek,bel,berg}. 

{\bf Advantages of the SME.}
The SME 
permits the identification 
and direct comparison 
of essentially all currently feasible experiments
searching for Lorentz and CPT violation. 
In addition, 
certain limits of the SME 
correspond to classical kinematics test models of relativity 
(such as the aforementioned Robertson's framework, 
its Mansouri-Sexl extension, 
or the $c^2$ model) \cite{km02}. 
Another advantage of the SME 
is the possibility of implementing 
additional desirable conditions 
besides coordinate independence. 
For instance, 
one can choose to require 
translational invariance, 
SU(3)$\times$SU(2)$\times$U(1) gauge symmetry, 
power-counting renormalizability, 
hermiticity,
and pointlike interactions. 
These conditions 
further restrict the parameter space for Lorentz breaking. 
One can also adopt simplifying choices, 
such as rotational invariance 
in certain coordinate systems. 
This latter assumption 
together with additional simplifications of the SME 
has been discussed in Ref.\ \cite{cg99}. 

\section{Varying couplings and the SME}
\label{vc}

In this section, 
we construct a classical cosmological solution 
in the framework of the pure $N=4$ supergravity 
in a four-dimensional spacetime. 
We show 
that this solution 
leads to a spacetime variation of both 
the fine-structure parameter $\al$ 
and the electromagnetic $\th$ angle. 
Such a model fails to be fully realistic in detail, 
but it is a limit of the $N=1$ supergravity in 11 dimensions,
which is contained in M-theory. 
It could therefore give insight 
into generic features 
expected to arise 
in a promising candidate underlying theory. 
We illustrate the associated Lorentz-violating effects 
by looking at the $\th$-angle variation, 
which gives rise to the ${(k_{AF})}^{\mu}$ term 
contained in the SME. 

{\bf Basics of the model.} 
In Planck units, 
the bosonic lagrangian 
for our $N=4$ supergravity in four dimensions 
takes the following form \cite{cj}: 
\beq
\cl=\sqrt{g}\left(-\frac 1 2 R
-\frac 1 4 M F_{\mu\nu} F^{\mu\nu}
-\frac 1 4 N F_{\mu\nu} \tilde{F}^{\mu\nu}
+\fr{{\prt_\mu A\prt^\mu A + \prt_\mu B\prt^\mu B}}{4B^2}\right). 
\label{lag2}
\eeq
Here, $g_{\mu\nu}$ represents the metric, 
and we have assumed 
that only one graviphoton, 
$F_{\mu\nu}$, 
is excited. 
The dual $\tilde{F}^{\mu\nu}=\ve^{\mu\nu\rh\si}F_{\rh\si}/2$ 
is defined as usual. 
The model also contains two scalars $A$ and $B$ 
that can be identified 
with an axion and a dilaton. 
The dependence of the couplings $M$ and $N$ 
on the scalars 
is fixed by the supergravity framework \cite{klp03}:
\beq
M=\fr
{B (A^2 + B^2 + 1)}
{(1+A^2 + B^2)^2 - 4 A^2},\quad
N=\fr
{A (A^2 + B^2 - 1)}
{(1+A^2 + B^2)^2 - 4 A^2}.
\label{N}
\eeq

In the pure $N=4$ supergravity in four dimensions, 
the graviphoton couples nonminimally to matter. 
Although the internal SO(4) symmetry can be gauged \cite{dfr77}, 
we adopt a phenomenological approach: 
in a realistic situation, 
the vector-matter interaction 
must be minimal. 
In what follows, 
we therefore can identify the graviphoton $F_{\mu\nu}$ 
with the electromagnetic field. 

{\bf Supergravity cosmology.} 
Next, 
we consider the above model in a cosmological context. 
We begin with the standard assumption 
of a homogeneous and isotropic universe. 
This implies 
that $F_{\mu\nu}\simeq 0$ on cosmological scales. 
We further take the universe 
to be flat, i.e., $k=0$. 
This is justified 
in light of recent measurements \cite{perlmutter,riess,garn}. 
The Friedmann-Robertson-Walker (FRW) line element 
has the conventional form: 
\beq
ds^2 = dt^2 - a^2(t) (dx^2 + dy^2 + dz^2) .
\label{frw}
\eeq
Here, the usual comoving coordinates have been adopted 
and $a(t)$ denotes the cosmological scale factor. 
As a consequence of the above assumptions, 
not only the scale factor, 
but also $A$ and $B$ depend on $t$ only. 

For phenomenological reasons 
it is necessary to model 
the known matter content of the universe. 
We employ a standard approach
and incorporate the energy-momentum tensor of dust, $T_{\mu\nu}$, 
into our framework. 
If $u^\mu$ is the unit timelike vector
orthogonal to the spatial hypersurfaces
and $\rh(t)$ is the energy density of the dust,
the usual arguments imply that $T_{\mu\nu} = \rh u_\mu u_\nu$. 
In the present model, 
this type of matter
arises from the fermionic sector
of our supergravity framework. 
At tree level, 
the scalars $A$ and $B$
do not couple to the fermions \cite{cj}, 
so that we can take $T_{\mu\nu}$ as conserved separately. 

It turns out 
that the equations of motion for our supergravity cosmology 
can be integrated analytically. 
For example, 
the time dependences of $A$ and $B$ 
are given by \cite{klp03}
\beq
A = \pm\la \tanh \left(\fr{1}{\ta} - \fr{1}{\ta_0}\right) + A_0,
\quad
B = \la \sech \left(\fr{1}{\ta} - \fr{1}{\ta_0}\right) ,
\label{be}
\eeq
where 
$\la$, $1/\ta_0$, and $A_0$ are integration constants.
The parameter time $\ta$ is defined by 
$\ta=\sqrt{3}/4\;{\rm arcoth}(\sqrt{3c_n/4c_1}\:t+1)$, 
which contains two more integration constants $c_n$ and $c_1$. 
The solution \rf{be} implies
that both $A$ and $B$ approach constant values 
at late times $t\rightarrow\infty$.
Thus, 
the values of the axion $A$ and the dilaton $B$ become fixed 
in our supergravity cosmology, 
despite the absence of a dilaton potential. 
This is essentially a consequence of energy conservation. 

{\bf Varying couplings.} 
Next, 
we consider excitations of $F_{\mu\nu}$ 
in the axion-dilaton background determined by Eq.\ \rf{be}. 
From a phenomenological viewpoint, 
we can take these excitations to be localized 
in spacetime regions small on cosmological scales. 
It is therefore appropriate 
to work in local inertial frames.

The conventional electrodynamics lagrangian 
in inertial coordinates can be taken as 
\beq
\cl_{\rm em} = 
-\fr{1}{4 e^2} F_{\mu\nu}F^{\mu\nu}
- \fr{\th}{16\pi^2} F_{\mu\nu} \tilde{F}^{\mu\nu}, 
\label{em}
\eeq
where we have allowed for a nontrivial $\th$-angle. 
The electromagnetic coupling is denoted by $e$. 
Comparison with our supergravity lagrangian \rf{lag2} shows 
that we can identify 
$e^2 \equiv 1/M$ and $\th \equiv 4\pi^2 N$. 
It is important to note 
that $M$ and $N$ 
are determined by the axion-dilaton background \rf{be}, 
so that $e$ and $\th$ become functions 
of the comoving time $t$. 
In arbitrary local inertial frames, 
the electromagnetic coupling and the $\th$ angle 
therefore exhibit related spacetime dependences. 

\begin{figure}
\begin{center}
\includegraphics[width=.8\textwidth]{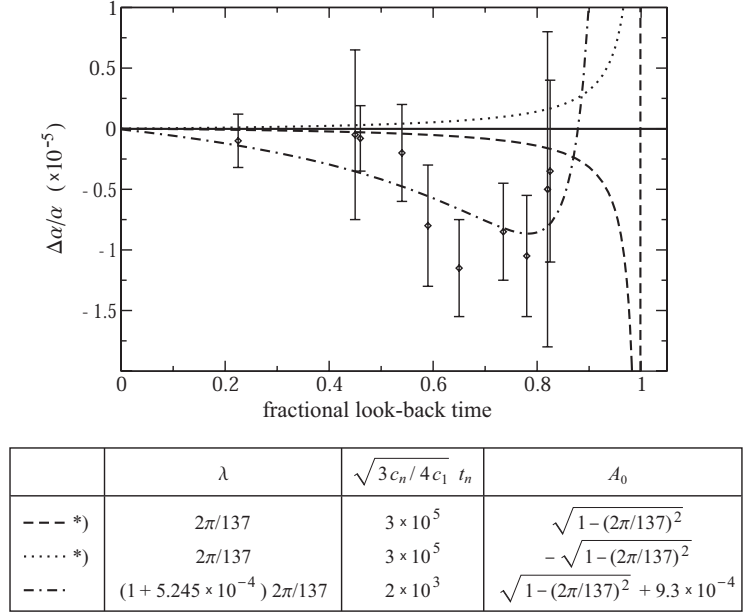}
\end{center}
\caption[]{
Relative change in the fine-structure parameter 
versus fractional look-back time to the big bang 
for various choices of integration constants \cite{03rl}. 
Note the qualitative differences and nonlinear features. 
Also shown are the Webb data \cite{webb}. 
}
\label{fig1}
\end{figure}

We remark in passing 
that the functional dependence 
of the fine-structure parameter $\al = e^2/4\pi$ 
on the comoving time 
can vary qualitatively 
with the choice of model parameters. 
Figure \rf{fig1} displays relative variations 
$\De\al/\al$
of the fine-structure parameter 
versus fractional look-back time $1-t/t_n$ 
to the big bang. 
Here, 
$t_n$ denotes the present age of the universe 
and $\De\al=\al(t)-\al(t_n)$. 
We have set the parameter $1/\ta_0$ to zero. 
The solid line corresponds to a constant $\al$. 
Each broken line 
represents a set of nontrivial choices for 
$\la$, $\sqrt{3c_n/4c_1}\:t_n$, and $A_0$. 
Input parameters leading to a variation 
consistent with the Oklo constraints \cite{damour96,fujii00,olive02} 
are labeled with an asterisk. 
The qualitative differences in the various plots, 
the nonlinear features, 
and the sign change for $\dot \al$ in the two cases with positive $A_0$ 
are apparent. 
Figure \rf{fig1} 
also depicts 
the recent experimental results \cite{webb} 
obtained from measurements of high-redshift spectra 
over periods of approximately $0.6t_n$ to $0.8t_n$ 
assuming $H_0=65$ km/s/Mpc and $(\Om_m ,\Om_\La)=(0.3,0.7)$. 

{\bf Lorentz violation.} 
The Lorentz violation associated with varying couplings 
becomes perhaps most transparent in the equations of motion. 
Incorporating charged matter 
described by a 4-current $j^{\nu}$, 
lagrangian \rf{lag2} yields in a local inertial frame: 
\beq
\fr{1}{e^2}\partial_{\mu}F^{\mu\nu}
-\fr{2}{e^3}(\partial_{\mu}e)F^{\mu\nu}
+\fr{1}{4\pi^2}(\partial_{\mu}\th)\tilde{F}^{\mu\nu}=j^{\nu} .
\label{Feom}
\eeq
Note that in the limit of spacetime-constant $e$ and $\th$, 
the conventional inhomogeneous Maxwell equations are recovered. 
In the axion-dilaton background \rf{be}, however,
the terms containing the gradients of $e$ and $\th$ 
are associated with apparent Lorentz violation: 
since the gradients can be treated 
as effectively nondynamical and constant 
on small cosmological scales, 
they select a preferred direction 
in the local inertial frame.
As a result, 
symmetry under boosts and rotations 
of electromagnetic fields is broken. 
Note that this type of Lorentz violation 
is not a feature 
of the particular coordinate system chosen. 
If a gradient is 
is nonzero in one local inertial frame 
associated with a small spacetime region, 
it is nonzero in {\it all} local inertial frames
associated with the region in question. 

By contrast, 
such a Lorentz-symmetry breakdown is absent 
in conventional FRW cosmologies 
that fail to contain spacetime-dependent scalars. 
Although global Lorentz invariance is usually violated, 
local Lorentz-symmetric inertial frames always exist. 
It is also important to note
that the above source 
for  Lorentz-violating effects 
is not a unique feature of our supergravity cosmology. 
Equation \rf{Feom} illustrates
that any smooth spacetime dependence 
of the couplings $e$ and $\th$
on cosmological scales leads to such effects. 
It is theoretically attractive 
to associate varying couplings 
with quantum scalar fields 
acquiring spacetime-dependent expectation values. 
However, 
from the perspective of Lorentz violation, 
classical scalars can be employed equally well. 
In fact, 
the variation of the coupling need not necessarily 
be driven by dynamical fields at all. 
This suggests 
that the above type of Lorentz breaking 
is a common feature of any model 
with spacetime-varying couplings. 

Next, 
we study 
how the effects of this mechanism 
fit into the framework of the SME 
and how our supergravity model 
helps to resolve conceptual issues 
in quantum field theories incorporating Lorentz breaking. 
This is best illustrated 
by considering the $\th$-angle term. 
An integration by parts 
yields an equivalent form 
of the electrodynamics lagrangian \rf{em} 
in a local inertial frame: 
\beq
\cl_{\rm em}^{\prime} = -\fr{1}{4 e^2}  F_{\mu\nu} F^{\mu\nu}
+\fr{1}{8\pi^2}(\prt_\mu\th) A_\nu \tilde{F}^{\mu\nu} .
\label{prlagr}
\eeq
The the last term 
on right-hand side of Eq.\ \rf{prlagr} 
gives a Chern-Simons-type contribution to the action. 
Such a term 
is contained in the minimal SME, 
and we can identify 
$(k_{AF})_\mu \equiv e^2 \prt_\mu\th/8\pi^2$. 
The presence of a nonzero $(k_{AF})_\mu$ 
in Eq.\ \rf{prlagr} 
demonstrates explicitly 
Lorentz and CPT breaking at the lagrangian level. 

The case of constant $e$ and $(k_{AF})_\mu$ 
has been discussed extensively 
in the literature \cite{cfj,ck98,jk,klink,adam}. 
Then, 
lagrangian \rf{prlagr} 
becomes translationally invariant 
resulting in an associated conserved energy 
that fails to be positive definite. 
This usually leads to instabilities in the theory, 
and the question arises 
how this problem is avoided in the present context 
of our positive-definite supergravity model.\footnote{The 
conserved symmetric energy-momentum tensor 
for the lagrangian \rf{lag2} 
acquires no contribution from the $N$ term 
because the latter is independent of $g^{\mu\nu}$. 
The other terms contributing to the energy density are positive definite.
} 

Although in most models 
a Chern-Simons-type term is assumed to arise 
in a fundamental theory, 
$(k_{AF})_\mu$ is typically treated 
as constant and nondynamical at low energies. 
In our supergravity cosmology, 
however, 
$(k_{AF})_\mu$ is associated 
with the dynamical scalars $A$ and $B$. 
Excitations of $F_{\mu\nu}$
therefore result in perturbations
$\de A$ and $\de B$ 
in the axion-dilaton background \rf{be}.
As a consequence, 
the energy-momentum tensor $(T^{\rm b})^{\mu\nu}$ 
of the background 
acquires an additional contribution, 
$(T^{\rm b})^{\mu\nu}\rightarrow 
(T^{\rm b})^{\mu\nu}+\de (T^{\rm b})^{\mu\nu}$. 
One can show \cite{klp03} 
that the contribution $\de (T^{\rm b})^{\mu\nu}$
does indeed compensate 
for the negative energies 
associated with a nonzero $(k_{AF})_\mu$. 

\section{Threshold analyses} 
\label{thres} 

Within the SME, 
it is straightforward to verify 
that Lorentz violation 
typically modifies one-particle dispersion relations 
\cite{cfj,ck97,ck98,kl01}. 
This feature 
permits the prediction of possible experimental signatures 
for Lorentz-symmetry breakdown 
based purely on kinematical arguments. 
For instance, 
primary ultrahigh-energy cosmic rays (UHECR) 
at energies up to eight orders of magnitude 
below the Planck scale have been observed. 
At such energies, 
Lorentz-violating effects 
might be pronounced 
relative to the ones in low-energy experiments. 
This leads to potentially observable 
threshold modifications for particle reactions, 
an idea 
that has been adopted in many recent studies of Lorentz-symmetry breakdown 
\cite{cg99,bert,jlm,koma,steck,jan03}. 
However, 
it is known \cite{rl03}
that some threshold analyses 
employ phenomenologically constructed dispersion relations 
that violate physics principles more fundamental than Lorentz symmetry. 

In this section, 
we investigate 
how some of the arbitrariness 
in the construction of dispersion-relation modifications 
can be removed. 
Our study 
relies on the principle of coordinate invariance 
and on the condition of compatibility 
with an effective dynamical framework like the SME. 
These two features appear fundamental enough 
for being physical requirements, 
while maintaining relative independence 
of the details of the Planck-scale theory. 
We also discuss causality and positivity, 
features 
that further add 
to the viability of threshold analyses. 
Throughout we assume exact conservation of energy and momentum. 

{\bf Coordinate-independent dispersion relations.} 
As argued in Sec.\ \ref{ci}, 
coordinate independence 
is essential in physics, 
despite the presence of Lorentz violation. 
In the published literature, 
the usual ansatz 
for modified dispersion relations 
is of the following form: 
\beq
{E}^2-{\vec{p}}^{\:2}=m^2+\de f(E,\vec{p})\; ,
\label{lvdr}
\eeq
where $m$ is the usual mass parameter 
and $p^{\mu}=(E,\vec{p})$ 
the 4-momentum. 
The function $\de f(E,\vec{p})$ 
controls the extent of the Lorentz violation. 
Coordinate independence 
requires $\de f$ to be a scalar, 
so that
\beq
\de f(E,\vec{p})=
\sum_{n\ge 1}\hspace{4mm}
\overbrace{\hspace{-3.5mm}T_{(n)}^{\;\al\be\;\cdots\;\;}}^{n\; \rm indices}
\hspace{-1mm}\underbrace{\hspace{.5mm}p_{\al}p_{\be}\;\cdots\;}
_{n\; \rm factors} \;\; .
\label{ansatz}
\eeq
Here, $T_{(n)}^{\;\al\be\;\cdots\;}$ denotes a constant tensor of rank $n$ 
representing the Lorentz-breaking background. 
The tensor indices $\al,\be,\,\ldots$ 
are distinct 
but each one is contracted with a 4-momentum factor. 
This ensures that all terms in the sum are observer Lorentz invariant. 
Under mild assumptions, 
Eq.\ \rf{ansatz} 
determines the most general Lorentz-violating dispersion relation 
compatible with coordinate independence. 

The implications of the general coordinate-independent ansatz \rf{ansatz} 
can be illustrated 
when the common assumption 
of rotation invariance in certain frames is made. 
In this case, 
the form of the Lorentz-violating tensor parameters $T_{(n)}$ 
is constrained by the imposed rotational symmetry. 
As a consequence, 
the correction $\de f$ fails to contain 
odd powers of the 3-momentum magnitude $|\vec{p}|$ \cite{rl03}. 
This result is to be contrasted 
with the common occurrence of $|\vec{p}|^3$ corrections 
in modified dispersion relations 
constructed by hand 
without reference to principles essential in physics. 

Note that a correction $\sim E |\vec{p}|^2$ 
is consistent with coordinate independence. 
Then, 
the question arises 
as to whether the usual ultrarelativistic relation 
$E\simeq|\vec{p}|$ 
can introduce an effective $|\vec{p}|^3$ modification. 
Although such a replacement 
may yield excellent approximations for the eigenenergies, 
it gives incorrect results in threshold analyses. 
This is intuitively reasonable 
because this replacement reintroduces 
the conventional degeneracy of the eigenenergies. 
An explicit example for the failure 
of the ultrarelativistic approximation 
is provided by photon decay 
into an electron-positron pair:
\beq
\ga\rightarrow e^++e^-.
\label{photodecay} 
\eeq
This process is kinematically forbidden in conventional physics. 
In the present Lorentz-violating context, 
the decay \rf{photodecay} is allowed 
when the coordinate-indepen{\-}dent correction $\sim E |\vec{p}|^2$ 
for photon, electron, and positron is used. 
However, 
when the correction term is approximated by $|\vec{p}|^3$, 
the process ceases to be kinematically permitted. 

{\bf Underlying dynamical framework.} 
The need for underlying dynamics 
can be illustrated with the following example. 
Consider the rotationally symmetric modified dispersion relation 
\beq
{(p^{\mu}p_{\mu}-m^2)}^2=|\vec{p}|^6.
\label{mdr} 
\eeq
Note that odd powers of $|\vec{p}|$ are absent 
compatible with coordinate independence. 
After the usual reinterpretation 
of the negative-energy solutions, 
the particle and antiparticle energies are given by 
\beq 
E_{\pm}^{(\al)}(\vec{p})=\sqrt{(-1)^{\al}
\fr{|\vec{p}|^{\,3}}{M}+\vec{p}^{\:2}+m^2}\; , 
\label{eigenen3} 
\eeq 
where $\al=1,2$ labels the two possible particle (antiparticle) energies, 
which perhaps correspond to different spin-type states. 
An analysis of the photon decay \rf{photodecay} 
employing these eigenenergies reveals 
that six kinematically distinct decays have to be considered. 
Note, however, 
that angular-momentum conservation associated 
with the enforced rotational invariance 
may preclude some of the six reactions. 
A proper study of this case 
therefore requires dynamical concepts. 

{\bf Causality and positivity.} 
Causality and energy positivity 
are fundamental requirements in physics. 
However, 
one or both of these requirements can be violated 
in the presence of Lorentz violation \cite{kl01}. 
From a conservative viewpoint, 
it is therefore natural to ask 
whether reaction-threshold kinematics is significantly affected 
when positivity and causality are imposed. 
Let $M$ and $m$ denote the scales 
of the fundamental theory and current low-energy physics, 
respectively. 
Then, 
the scale $p_{s{\textrm -}c}$ for the occurrence of spacelike momenta 
(and thus negative energies in certain frames) 
or causality problems 
can be as low as \cite{kl01}
\beq
p_{s{\textrm -}c}\sim{\cal O}(\sqrt{mM}\:)\; .
\label{scale}
\eeq
For example, 
if $m$ is the proton mass 
and $M$ is taken to be the Planck scale, 
then $p_{s{\textrm -}c}\sim3\times 10^{18}$ eV. 
UHECRs with a spectrum extending beyond $10^{20}$ eV 
are often employed to bound Lorentz breaking 
or to suggest evidence for Lorentz violation. 
Thus, imposing causality and positivity 
could require modifications in threshold analyses. 

As a specific example, 
consider the decay  
\beq
\ga\rightarrow \pi^0+\ga, 
\label{decay} 
\eeq
which is kinematically forbidden in conventional physics. 
The usual  dispersion-relation modifications 
in the literature permitting this process 
are associated with causality or positivity violations: 
take the photon energy $E$ to be given by  
$E=({\vec{p}}^{\: 2}+\de f(\vec{p}))^{1/2}$,  
where $\de f(\vec{p})$ 
excludes a mass term \footnote{A mass term 
would yield a Lorentz-invariant 
(but gauge-symmetry violating) contribution, 
which is not of interest in the present context.} 
and depends only on the photon 3-momentum $\vec{p}$. 
Then, 
the point $(E,\vec{p})=(0,\vec{0})$ on the momentum-space lightcone 
must satisfy the modified dispersion relation. 
For some 3-momenta $\vec{p}\neq\vec{0}$, 
a nontrivial correction $\de f(\vec{p})$ forces $E(\vec{p})$ 
to curve to the outside or inside of the lightcone 
leading to spacelike 4-momenta or superluminal group velocities, 
respectively. 
If, 
however, 
the photon dispersion relation is taken to be the conventional one, 
spacelike pions are required for the decay to occur.\footnote{The 
allowed phase space for the decay products 
in the case of a lightlike pion 4-momentum 
is a set of measure zero 
leading at best to a suppressed rate for the reaction. 
} 

\section{Conclusion} 
\label{conc} 

Although Lorentz and CPT invariance 
are deeply ingrained 
in the currently accepted laws of physics, 
there are a variety of candidate underlying theories 
admitting the violation of these symmetries. 
The ultrahigh sensitivity of many Lorentz and CPT tests 
therefore permits the experimental search for Planck-scale physics. 

Spacetime-dependent couplings 
are one potential source 
for apparent Lorentz and CPT violation: 
the gradient of such couplings 
in the equations of motion 
selects a preferred spacetime direction 
in the effective vacuum. 
We have argued 
that variations of couplings 
are natural in a cosmological context 
of candidate fundamental theories. 

The first-order Lorentz-violating effects 
resulting from varying couplings 
and other mechanisms for Lorentz-symmetry breakdown 
are described by the SME. 
At the level of effective quantum field theory, 
the SME 
is the most general dynamical framework 
for Lorentz and CPT violation 
that is compatible 
with the fundamental principle of coordinate independence. 

Threshold analyses with modified dispersion relations 
are conceptually clean Lorentz tests 
and are best performed within the SME. 
Many purely kinematical threshold considerations in the literature 
are insufficient for bounding Lorentz breaking 
because they violate coordinate independence 
or other fundamental principles. 

\section*{Acknowledgments} 
The author wishes to thank H.V.\ Klapdor-Kleingrothaus 
for the invitation 
to the Beyond '03 Meeting. 
This work was supported in part 
by the Centro Multidisciplinar de Astrof\'{\i}sica (CENTRA) 
and by the Funda\c{c}\~ao para a Ci\^encia e a Tecnologia (Portugal) 
under grant POCTI/FNU/49529/2002.

%


\begin{thebibliography}{xx}
\addcontentsline{toc}{section}{References}

\bibitem{klp03}
V.A.\ Kosteleck\'y, R.\ Lehnert, M.J.\ Perry: 
Phys.\ Rev.\ D {\bf 68}, 123511 (2003)

\bibitem{ck97} 
D.\ Colladay, V.A.\ Kosteleck\'y: 
Phys.\ Rev.\ D {\bf 55}, 6760 (1997) 

\bibitem{ck98} 
D.\ Colladay, V.A.\ Kosteleck\'y:  
Phys.\ Rev.\ D {\bf 58}, 116002 (1998) 

\bibitem{jk}
R.\ Jackiw, V.A.\ Kosteleck\'y: 
Phys.\ Rev.\ Lett.\ {\bf 82}, 3572 (1999) 
 
\bibitem{kl01} 
V.A.\ Kosteleck\'y, R.\ Lehnert: 
Phys.\ Rev.\ D {\bf 63}, 065008 (2001) 

\bibitem{klp02}
V.A.\ Kosteleck\'y, C.D.\ Lane, A.G.M.\ Pickering: 
Phys.\ Rev.\ D {\bf 65}, 056006 (2002) 

\bibitem{grav} 
V.A.\ Kosteleck\'y: 
hep-th/0312310 

\bibitem{ktev} 
KTeV Collaboration, 
H.\ Nguyen: 
`CPT Results from KTeV'. 
In: {\it CPT and Lorentz Symmetry II}, 
ed.\ by V.A.\ Kosteleck\'y 
(World Scientific, Singapore, 2002) 

\bibitem{opal} 
OPAL Collaboration, 
R.\ Ackerstaff 
{\it et al.}: 
Z.\ Phys.\ C {\bf 76}, 401 (1997) 

\bibitem{delphi} 
DELPHI Collaboration, 
M.\ Feindt 
{\it et al.}: 
preprint DELPHI 97-98 CONF 80 (1997) 

\bibitem{belle} 
BELLE Collaboration,
K.\ Abe 
{\it et al.}: 
Phys.\ Rev.\ Lett.\ {\bf 86}, 3228 (2001) 

\bibitem{focus} 
FOCUS Collaboration, 
J.M.\ Link {\it et al.}: 
Phys.\ Lett.\ B {\bf 556}, 7 (2003) 

\bibitem{ck95a} 
D.\ Colladay, V.A.\ Kosteleck\'y: 
Phys.\ Lett.\ B {\bf 344}, 259 (1995) 

\bibitem{ck95b} 
D.\ Colladay, V.A.\ Kosteleck\'y: 
Phys.\ Rev.\ D {\bf 52}, 6224 (1995) 

\bibitem{ck95c} 
D.\ Colladay, V.A.\ Kosteleck\'y: 
Phys.\ Lett.\ B {\bf 511}, 209 (2001) 

\bibitem{kv96} 
V.A.\ Kosteleck\'y, R.\ Van Kooten: 
Phys.\ Rev.\ D {\bf 54}, 5585 (1996) 

\bibitem{baryo} 
O.\ Bertolami, D.\ Colladay, V.A.\ Kosteleck\'y, R.\ Potting: 
Phys.\ Lett.\ B {\bf 395}, 178 (1997) 

\bibitem{k98} 
V.A.\ Kosteleck\'y: 
Phys.\ Rev.\ Lett.\ {\bf 80}, 1818 (1998) 

\bibitem{k99} 
V.A.\ Kosteleck\'y: 
Phys.\ Rev.\ D {\bf 61}, 016002 (2000) 

\bibitem{k01} 
V.A.\ Kosteleck\'y: 
Phys.\ Rev.\ D {\bf 64}, 076001 (2001) 

\bibitem{isg01} 
N.\ Isgur \etal: 
Phys.\ Lett.\ B {\bf 515}, 333 (2001) 

\bibitem{hu99} 
L.R.\ Hunter 
{\it et al.}: 
`Limits on Local Lorentz Invariance 
from Hg and Cs Magnetometers'. 
In: {\it CPT and Lorentz Symmetry}, 
ed.\ by V.A.\ Kosteleck\'y 
(World Scientific, Singapore, 1999) 

\bibitem{bear00} 
D.\ Bear 
{\it et al.}: 
Phys.\ Rev.\ Lett.\ {\bf 85}, 5038 (2000) 

\bibitem{hum00} 
M.A.\ Humphrey 
{\it et al.}: 
Phys.\ Rev.\ A {\bf 62}, 063405 (2000) 

\bibitem{kl99a} 
V.A.\ Kosteleck\'y, C.D.\ Lane: 
Phys.\ Rev.\ D {\bf 60}, 116010 (1999) 

\bibitem{kl99b} 
V.A.\ Kosteleck\'y, C.D.\ Lane: 
J.\ Math.\ Phys.\ {\bf 40}, 6245 (1999) 

\bibitem{blu02} 
R.\ Bluhm \etal: 
Phys.\ Rev.\ Lett.\ {\bf 88}, 090801 (2002) 

\bibitem{blu03} 
R.\ Bluhm \etal: 
hep-ph/0306190 

\bibitem{sudar02} 
D.\ Sudarsky, L.\ Urrutia, H.\ Vucetich: 
Phys.\ Rev.\ Lett.\ {\bf 89}, 231301 (2002) 

\bibitem{phil01} 
D.F.\ Phillips
{\it et al.}: 
Phys.\ Rev.\ D {\bf 63}, 111101 (2001) 

\bibitem{pnthhh} 
R.\ Bluhm \etal: 
Phys.\ Rev.\ Lett.\ {\bf 82}, 2254 (1999) 

\bibitem{deh99} 
H.\ Dehmelt 
{\it et al.}: 
Phys.\ Rev.\ Lett.\ {\bf 83}, 4694 (1999) 

\bibitem{mit99} 
R.\ Mittleman 
{\it et al.}: 
Phys.\ Rev.\ Lett.\ {\bf 83}, 2116 (1999) 

\bibitem{gab99} 
G.\ Gabrielse 
{\it et al.}: 
Phys.\ Rev.\ Lett.\ {\bf 82}, 3198 (1999) 

\bibitem{bluhm97} 
R.\ Bluhm \etal: 
Phys.\ Rev.\ Lett.\ {\bf 79}, 1432 (1997) 

\bibitem{bluhm98} 
R.\ Bluhm \etal: 
Phys.\ Rev.\ D {\bf 57}, 3932 (1998) 

\bibitem{bluhm00} 
R.\ Bluhm, V.A.\ Kosteleck\'y: 
Phys.\ Rev.\ Lett.\ {\bf 84}, 1381 (2000) 

\bibitem{heck02} 
B.\ Heckel: 
`Testing CPT and Lorentz Symmetry 
with a Spin-Polarized Torsion Pendulum'. 
In: {\it CPT and Lorentz Symmetry II}, 
ed.\ by V.A.\ Kosteleck\'y 
(World Scientific, Singapore, 2002) 

\bibitem{iltan2} 
E.O.\ Iltan: 
hep-ph/0308151 

\bibitem{iltan3} 
E.O.\ Iltan: 
hep-ph/0309154 

\bibitem{rl04} 
R.\ Lehnert: 
hep-ph/0401084 

\bibitem{cfj} 
S.\ Carroll, G.\ Field, R.\ Jackiw: 
Phys.\ Rev.\ D {\bf 41}, 1231 (1990) 

\bibitem{km} 
V.A.\ Kosteleck\'y, M.\ Mewes: 
Phys.\ Rev.\ Lett.\ {\bf 87}, 251304 (2001) 

\bibitem{km02} 
V.A.\ Kosteleck\'y, M.\ Mewes: 
Phys.\ Rev.\ D {\bf 66}, 056005 (2002) 

\bibitem{brax} 
H.\ M\"uller \etal: 
Phys.\ Rev.\ D {\bf 67}, 056006 (2003) 

\bibitem{lipa} 
J.A.\ Lipa \etal: 
Phys.\ Rev.\ Lett.\ {\bf 90}, 060403 (2003) 

\bibitem{sh03}
G.M.\ Shore: 
Contemp.\ Phys.\ {\bf 44}, 503 {2003} 

\bibitem{muons1} 
V.W.\ Hughes
{\it et al.}: 
Phys.\ Rev.\ Lett.\ {\bf 87}, 111804 (2001) 

\bibitem{muons2} 
R.\ Bluhm \etal: 
Phys.\ Rev.\ Lett.\ {\bf 84}, 1098 (2000) 

\bibitem{iltan1} 
E.O.\ Iltan: 
JHEP {\bf 0306}, 016 (2003). 

\bibitem{cg99} 
S.\ Coleman, S.L.\ Glashow: 
Phys.\ Rev.\ D {\bf 59}, 116008 (1999) 

\bibitem{neutrinos}
V.\ Barger, S.\ Pakvasa, T.\ Weiler, K.\ Whisnant: 
Phys.\ Rev.\ Lett.\ {\bf 85}, 5055 (2000) 

\bibitem{neu1} 
V.A.\ Kosteleck\'y, M.\ Mewes: 
hep-ph/0308300 

\bibitem{neu2} 
V.A.\ Kosteleck\'y, M.\ Mewes: 
Phys.\ Rev.\ D, in press [hep-ph/0309025] 

\bibitem{rl03} 
R.\ Lehnert: 
Phys.\ Rev.\ D {\bf 68}, 085003 (2003)

\bibitem{baren01} 
G.\ Barenboim \etal: 
JHEP {\bf 0210}, 001 (2002)

\bibitem{baren02} 
G.\ Barenboim, J.\ Lykken: 
Phys.\ Lett.\ B {\bf 554}, 73 (2003) 

\bibitem{green02} 
O.W.\ Greenberg: 
Phys.\ Rev.\ Lett.\ {\bf 89}, 231602 (2002) 

\bibitem{green03} 
O.W.\ Greenberg: 
Phys.\ Lett.\ B {\bf 567}, 179 (2003) 

\bibitem{dsr1} 
G.\ Amelino-Camelia: 
Phys.\ Lett.\ B {\bf 510}, 255 (2001) 

\bibitem{dsr2} 
J.\ Magueijo, L.\ Smolin: 
Phys.\ Rev.\ Lett.\ {\bf 88}, 190403 (2002) 

\bibitem{luk03} 
J.\ Lukierski, A.\ Nowicki: 
Int.\ J.\ Mod.\ Phys.\ A {\bf 18}, 7 (2003) 

\bibitem{rem03} 
J.\ Rembielinski, K.A.\ Smolinski: 
Bull.\ Soc.\ Sci.\ Lett.\ {\L}\'od\'z S\'er.\ Rech.\ D\'eform.\
{\bf 53}, 57 (2003) 

\bibitem{ahl02} 
D.V.\ Ahluwalia: 
gr-qc/0212128 

\bibitem{gru03} 
D.\ Grumiller, W.\ Kummer, D.V.\ Vassilevich: 
Ukr.\ J.\ Phys.\ {\bf 48}, 329 (2003) 

\bibitem{schu03} 
R.\ Sch\"utzhold, W.G.\ Unruh: 
gr-qc/0308049 

\bibitem{ks89a}
V.A.\ Kosteleck\'y, S.\ Samuel: 
Phys.\ Rev.\ D {\bf 39}, 683 (1989) 

\bibitem{ks89b}
V.A.\ Kosteleck\'y, S.\ Samuel: 
Phys.\ Rev.\ D {\bf 40}, 1886 (1989) 

\bibitem{ks89c}
V.A.\ Kosteleck\'y, S.\ Samuel: 
Phys.\ Rev.\ Lett.\ {\bf 63}, 224 (1989) 

\bibitem{ks91}
V.A.\ Kosteleck\'y, S.\ Samuel: 
Phys.\ Rev.\ Lett.\ {\bf 66}, 1811 (1991) 

\bibitem{kp91} 
V.A.\ Kosteleck\'y, R.\ Potting: 
Nucl.\ Phys.\ B {\bf 359}, 545 (1991) 

\bibitem{kp96} 
V.A.\ Kosteleck\'y, R.\ Potting: 
Phys.\ Lett.\ B {\bf 381}, 89 (1996) 

\bibitem{kpp00} 
V.A.\ Kosteleck\'y, M.\ Perry, R.\ Potting: 
Phys.\ Rev.\ Lett.\ {\bf 84}, 4541 (2000) 

\bibitem{kp01} 
V.A.\ Kosteleck\'y, R.\ Potting: 
Phys.\ Rev.\ D {\bf 63}, 046007 (2001) 

\bibitem{ell98} 
N.E.\ Mavromatos:  
these proceedings 

\bibitem{suv} 
D.\ Sudarsky, L.\ Urrutia, H.\ Vucetich: 
Phys.\ Rev.\ D {\bf 68}, 024010 (2003) 

\bibitem{klink} 
F.R.\ Klinkhamer: 
Nucl.\ Phys.\ B {\bf 578}, 277 (2000) 

\bibitem{amu00} 
J.\ Alfaro, H.A.\ Morales-T\'ecotl, L.F.\ Urrutia: 
Phys.\ Rev.\ Lett.\ {\bf 84}, 2318 (2000) 

\bibitem{amu02} 
J.\ Alfaro, H.A.\ Morales-T\'ecotl, L.F.\ Urrutia: 
Phys.\ Rev.\ D {\bf 65}, 103509 (2002) 

\bibitem{chklo} 
S.M.\ Carroll \etal: 
Phys.\ Rev.\ Lett.\ {\bf 87}, 141601 (2001) 

\bibitem{gur} 
Z.\ Guralnik \etal: 
Phys.\ Lett.\ B {\bf 517}, 450 (2001) 

\bibitem{ani} 
A.\ Anisimov \etal: Phys.\ Rev.\ D {\bf 68}, 085003 (2003)
hep-ph/0106356 

\bibitem{carl} 
C.E.\ Carlson \etal: 
Phys.\ Lett.\ B {\bf 518}, 201 (2001) 

\bibitem{blpr} 
O.\ Bertolami \etal: 
Phys.\ Rev.\ D, in press [astro-ph/0310344]

\bibitem{bek} 
M.S.\ Berger and V.A.\ Kosteleck\'y: 
Phys.\ Rev.\ D {\bf 65}, 091701(R) (2002) 

\bibitem{bel} 
H.\ Belich \etal: 
Phys.\ Rev.\ D {\bf 68}, 065030 (2003) 

\bibitem{berg} 
M.S.\ Berger: 
Phys.\ Rev.\ D {\bf 68}, 115005 (2003) 

\bibitem{cj} 
E.\ Cremmer, B.\ Julia: 
Nucl.\ Phys.\ B {\bf 159}, 141 (1979) 

\bibitem{dfr77} 
A.\ Das, M.\ Fischler, M.\ Ro\v{c}ek: 
Phys.\ Rev.\ D {\bf 16}, 3427 (1977) 

\bibitem{perlmutter} 
S.J.\ Perlmutter  \etal, 
[Supernova Cosmology Project Collaboration]: 
Nature {\bf 391}, 51 (1998) 

\bibitem{riess} 
A.G.\  Riess  \etal, 
[Supernova Search Team Collaboration]: 
Astron.\ J.\ {\bf 116}, 1009 (1998) 

\bibitem{garn} 
P.M.\ Garnavich \etal: 
Astrophys.\ J.\  {\bf 509}, 74 (1998) 

\bibitem{03rl} 
R.\ Lehnert: 
hep-ph/0308208 

\bibitem{webb} 
J.K.\ Webb \etal: 
Phys.\ Rev.\ Lett.\ {\bf 87}, 091301 (2001)
and references therein 

\bibitem{damour96}
T.\ Damour, F.\ Dyson: 
Nucl.\ Phys.\ B {\bf 480}, 37 (1996) 

\bibitem{fujii00} 
Y.\ Fujii \etal: 
Nucl.\ Phys.\ B {\bf 573}, 377 (2000) 

\bibitem{olive02} 
K.\ Olive \etal: 
hep-ph/0205269 

\bibitem{adam} 
C.\ Adam, F.R.\ Klinkhamer: 
Nucl.\ Phys.\ B {\bf 607}, 247 (2001) 

\bibitem{bert} 
O.\ Bertolami, C.S.\ Carvalho: 
Phys.\ Rev.\ D {\bf 61}, 103002 (2000) 

\bibitem{jlm} 
T.\ Jacobson, S.\ Liberati, D.\ Mattingly: 
Phys.\ Rev.\ D {\bf 66}, 081302 (2002) 

\bibitem{koma}
T.J.\ Konopka, S.A.\ Major: 
New J.\ Phys.\ {\bf 4}, 57 (2002) 

\bibitem{steck} 
F.W.\ Stecker: 
astro-ph/0308214 

\bibitem{jan03} 
M.\ Jankiewicz \etal: 
hep-ph/0312221 

\end{thebibliography}
\end{document}